\documentclass[12pt]{article}
\usepackage{latexsym}
\usepackage{epsfig,amssymb,euscript,slashed}
\usepackage[linktocpage]{hyperref}
\usepackage{amsmath}
\usepackage{cite} 
\usepackage{array,calc,epsfig}
\usepackage{bbm}
\usepackage{fancybox}

\numberwithin{equation}{section} %%
\usepackage[]{graphicx}
\usepackage{color}

\begin{document}
\font\cmss=cmss10 \font\cmsss=cmss10 at 7pt

\begin{flushright}{
%\scriptsize DFPD-17-TH-xx \\  
\scriptsize QMUL-PH-20-10}
\end{flushright}
\hfill
\vspace{14pt}
\begin{center}
{\Large 
\textbf{The CFT$_6$ origin of all tree-level\\ 4-point correlators in AdS$_3 \times S^3$
}}
\end{center}

\vspace{8pt}
\begin{center}
{\textsl{Stefano Giusto$^{\,1, 2}$, Rodolfo Russo$^{\,3}$, Alexander Tyukov$^{\,1, 2}$ and Congkao Wen$^{\,3}$}}

\vspace{1cm}

\textit{\small ${}^{1}$ Dipartimento di Fisica ed Astronomia ``Galileo Galilei",  Universit\`a di Padova,\\Via Marzolo 8, 35131 Padova, Italy} \\  \vspace{6pt}

\textit{\small ${}^{2}$ I.N.F.N. Sezione di Padova,
Via Marzolo 8, 35131 Padova, Italy}\\
\vspace{6pt}

\textit{\small ${}^{3}$ Centre for Research in String Theory, School of Physics and Astronomy\\
Queen Mary University of London,
Mile End Road, London, E1 4NS,
United Kingdom}\\
\vspace{6pt}

\end{center}

\vspace{12pt}

\begin{center}
\textbf{Abstract}
\end{center}

\vspace{4pt} {\small
  \noindent
 We provide strong evidence that all tree-level 4-point holographic correlators in AdS$_3 \times S^3$ are constrained by a hidden 6D conformal symmetry. This property has been discovered in the AdS$_5 \times S^5$ context and noticed in the tensor multiplet subsector of the AdS$_3 \times S^3$ theory. Here we extend it to general AdS$_3 \times S^3$ correlators which contain also the chiral primary operators of spin zero and one that sit in the gravity multiplet. The key observation is that the 6D conformal primary field associated with these operators is not a scalar but a  self-dual $3$-form primary.  As an example, we focus on the correlators involving two fields in the tensor multiplets and two in the gravity multiplet and show that all such correlators are encoded in a conformal 6D correlator between two scalars and two self-dual $3$-forms, which is determined by three functions of the cross ratios. We fix these three functions by comparing with the results of the simplest correlators derived from an explicit supergravity calculation.

\vspace{1cm}

\thispagestyle{empty}

\vfill
\vskip 5.mm
\hrule width 5.cm
\vskip 2.mm
{
\noindent  {\scriptsize e-mails:  {\tt stefano.giusto@pd.infn.it, r.russo@qmul.ac.uk, tyukov@pd.infn.it, c.wen@qmul.ac.uk} }
}

\setcounter{footnote}{0}
\setcounter{page}{0}

\newpage

%%%%%%%%%%%%%%%%%%%%%%%%%%%%%%%%%%%%%%%%%

\section{Introduction}

Several different approaches have been developed to study the correlators of local operators in holographic CFTs when it is possible to exploit the dual description in terms of a weakly coupled gravity theory. The traditional approach uses Witten's diagrams in AdS~\cite{Witten:1998qj} which has more recently been complemented by new tools such as the Mellin space formulation~\cite{Penedones:2010ue,Fitzpatrick:2011ia}, the ``position space'' method developed in~\cite{Rastelli:2016nze,Rastelli:2017udc}, the use of large spin perturbation theory~\cite{Alday:2016njk} and Lorentzian inversion formula~\cite{Caron-Huot:2017vep}\cite{Alday:2017vkk}, and the approach based on microstate geometries of~\cite{Galliani:2017jlg,Bombini:2017sge}. In the case of ${\cal N}=4$ SYM/ AdS$_5 \times S^5$ supergravity, these new techniques made it possible to study explicitly a large class of correlators and to extracting interest CFT data such as couplings and anomalous dimensions~\cite{Aprile:2017xsp,Aprile:2018efk}. This has led to a remarkable observation~\cite{Caron-Huot:2018kta}: the tree-level  4-point  supergravity amplitudes in AdS$_5 \times S^5 $ enjoy a 10D hidden conformal symmetry and this can be used to write compact recursion relations capturing all the tree-level holographic correlators of four half-BPS operators in ${\cal N}=4$ SYM.

It has been noticed~\cite{Rastelli:2019gtj, Giusto:2019pxc} that the holographic 4-point  correlators in AdS$_3\times S^3$ duality share some key properties with the AdS$_5$ cousin and so it is natural to ask whether a hidden conformal symmetry is present also in this case. The aim of this letter is to answer in an affirmative way this question and to show how to derive the recursion relations capturing all AdS$_3$ holographic correlators in the tree-level supergravity approximation. One aspect that makes this question interesting is that it is in general difficult to apply some of the modern techniques to holographic dualities involving a CFT$_2$. The chiral nature of the theory implies that the results for the 4-point correlators can depend separately on the cross-ratio $z$ and $\bar{z}$ so it is not known in general how to rewrite the results in Mellin space. Furthermore the CFT$_2$ considered here is a ${\cal N}=(4,4)$ SCFT and so has only half the amount of supercharges with respect to the AdS$_5$ case. SCFTs of this type were discussed in~\cite{Zhou:2018ofp}, but the Mellin bootstrap approach adopted there cannot be directly applied to this case for the reason mentioned above.

However there are indications suggesting that the pattern discovered in~\cite{Caron-Huot:2018kta} should be at play also in the AdS$_3\times S^3$ case. First, all the 4-point ${\cal N}=(2,0)$ 6D supergravity amplitudes~\cite{Heydeman:2018dje} relevant for the flat-space limit enjoy a hidden 6D conformal symmetry. Then, when focusing just on external states that are ``matter'' multiplets (i.e. tensor multiplet of the ${\cal N}=(2,0)$ 6D supergravity), it was shown~\cite{Rastelli:2019gtj} that all 4-point  holographic correlators derived in~\cite{Rastelli:2019gtj,Giusto:2019pxc} can be obtained via a recursion relation from the lowest AdS$_3$/CFT$_2$ 4-point correlator obtained in~\cite{Giusto:2018ovt}. We will first review these aspects and then show that the approach of~\cite{Giusto:2018ovt,Giusto:2019pxc} provides a natural interpretation of the known examples and a concrete way to construct a complete implementation of the 6D hidden conformal symmetry for all the multiplets in the theory. The crucial observation is that the 6D conformal field associated with the gravity multiplet operators is a self-dual 3-form, instead of a scalar as for the tensor multiplets or the $AdS_5$ case. As an example, we will work out how the hidden conformal symmetry constrains the correlators with two fields in the matter multiplets and two fields in the gravity multiplet. Of course in order to obtain an explicit recursion relation one needs also the results for some correlators which fix the initial data of the recursion.  We obtained these correlators by generalising the approach of~\cite{Giusto:2019pxc}; here we will quote just the results we need and refer to a forthcoming paper~\cite{GRTW} for their derivation. The framework presented in this work should make it possible to bring our knowledge of holographic correlators in AdS$_3\times S^3$ up to the same level as the AdS$_5 \times S^5$ counterpart and start a systematic study of the OPE data in the gravity regime, an analysis of loop corrections and possibly also of string corrections by adapting to AdS$_3$ successful approaches in the AdS$_5$ case~\cite{Alday:2017xua, Aprile:2017bgs, Aprile:2017qoy, Alday:2018pdi, Alday:2018kkw, Drummond:2019odu, Aprile:2019rep, Alday:2019nin, Drummond:2020dwr, Binder:2019jwn, Chester:2019pvm, Chester:2019jas, Chester:2020dja}. 

\section{Hints of a hidden 6D conformal symmetry}
\label{sec:hints}

Let us start from the tree-level 4-point  amplitude in ${\cal N}=(2,0)$ supergravity in flat space~\cite{Heydeman:2018dje}
\begin{equation}
  \label{eq:fs4p}
  \mathcal{A}_4 = G_6 \delta^8(Q) \delta^6(P) \frac{[1_{\hat a_1} 2_{\hat a_2} 3_{\hat a_3} 4_{\hat a_4}][1_{\hat b_1} 2_{\hat b_2} 3_{\hat b_3} 4_{\hat b_4}] }{s_{12} s_{23} s_{13}}\;,
\end{equation}
where $G_6$ is the 6D Newton constant, $s_{ij}=(p_i+p_j)^2$ are the Mandelstam variables, $\delta^6(P)$ indicates the standard momentum conservation, and the remaining ingredients are written in terms of 6D spinor helicity formalism: $p_{i\,\mu} (\Gamma^{\mu})^{AB} = \lambda^A_{i\,a} \lambda_{i}^{B, a} =  {1\over 2} \epsilon^{ABCD}  \tilde{\lambda}_{i\,, C\, \hat{a}} \tilde{\lambda}_{i\, D}^{\hat a}$.   The index $i$ indicates the external particle, $\mu$ is a vector index and $A,B,\ldots$ are spinor indices of 6D Lorentz group, and $a$ and $\hat{a}$ are $SU(2) \times SU(2)$ indices labelling the little group $SO(4)$. Finally $\delta^8(Q)$ involves the supercharges and scales as $\lambda^8$, while the square parenthesis in~\eqref{eq:fs4p} is defined as $[i_{\hat{a}_1} j_{\hat a_2} k_{\hat a_3} l_{\hat a_4}]:= \epsilon^{ABCD} \tilde{\lambda}_{i \, A\, \hat{a}_1}  \tilde{\lambda}_{j \, B\, \hat{a}_2}  \tilde{\lambda}_{k \, C\, \hat{a}_3} \tilde{\lambda}_{l \, D\, \hat{a}_4}$.

As its 10D counterpart, this amplitude enjoys some special features. The combination $G_6 \delta^8(Q)$ is dimensionless and we will focus on the truncated amplitude $\tilde{A}_4$ that does not contain this factor. By writing the 6D conformal generators in terms of spinor helicity variables~\cite{Huang:2010rn} 
\begin{equation}
  \label{eq:DKgen}
  D = {1 \over 2} \sum_i \left( \tilde{\lambda}_{i\, A\, \hat{a} }  {\partial \over \partial \tilde{\lambda}_{i\, A\, \hat{a}} }   + 4 \right)\;,~~~
K^{A B} = \sum_i  { \partial^2 \over \partial \tilde{\lambda}_{i\, \hat{a}\, A} \partial  \tilde{\lambda}^{\hat a}_{i\, B} } \;,
\end{equation}
it is possible to check explicitly that $\tilde{A}_4$ is annihilated by both $D$ and $K$. Let us conclude this discussion of the flat space amplitude, by pointing out that it is easy to separate the matter and the gravity parts in~\eqref{eq:fs4p}: in order to select a particle in the gravity multiplet for the $i^{\rm th}$ external state one needs to symmetrise the little group indices $\hat{a}_i$ and $\hat{b}_i$, while, if the indices are contracted with $\epsilon^{\hat{a}_i \hat{b}_i}$, then a particle in a tensor multiplet is selected. If all external states are taken to have antisymmetric little group indices, \eqref{eq:fs4p} simplifies yielding\footnote{We generalised the result obtained from~\eqref{eq:fs4p} to the case of particles in different tensor multiplets: each contribution $1/s_{ij}$ is multiplied by delta functions ensuring that in the $s_{ij}$ the particles $i$, $j$ involved are in the same tensor multiplet.}
\begin{equation}
  \label{eq:Aten}
\tilde{A}_{\rm ten} \sim \delta^{6}(P)\left(\frac{\delta_{f_1 f_2} \delta_{f_3 f_4}}{s_{12}}+\frac{\delta_{f_1 f_4} \delta_{f_2 f_3}}{s_{23}}+\frac{\delta_{f_1 f_3} \delta_{f_2 f_4}}{s_{13}}\right)\;.
\end{equation}  

When considered in the AdS$_3 \times S^3$ background the ${\cal N}=(2,0)$ supergravity discussed above captures the strong coupling limit of a ${\cal N}=(4,4)$ SCFT$_2$. Let us recall the main features of this theory. The R-symmetry group $SU(2)_L\times SU(2)_R$ can be identified with the isometries of the $S^3$ on the bulk side. The Chiral Primary Operators (CPOs) of the theory are labelled by the holomorphic and antihomorphic conformal dimensions $(h,\bar{h})$ and are in the $(j,\bar{j})=(h,\bar{h})$ representation of the R-symmetry. For each tensor multiplet there is a family of CPOs $s_k$ with the quantum numbers $(h,\bar{h})=(k/2,k/2)$ with $k=1,2,\ldots$. There is another left/right symmetric family of CPOs $\sigma_k$ with $(h,\bar{h})=(k/2,k/2)$ and $k=2,3,\ldots$. Finally there are two families of CPOs $V^\pm_k$ with $(h,\bar{h})=(k/2,k/2+1)$ and $(h,\bar{h})=(k/2+1,k/2)$ respectively and\footnote{For $k=0$ these CPO represent the R-symmetry currents and their correlators are determined by the affine Ward identities in terms of lower points correlators. We will not consider them in our analysis.} $k=0,1,\ldots$. Here we are following the notation of~\cite{Rastelli:2019gtj}, see the Tables 1--3 in that reference for more details. From the 6D point of view, the CPOs $\sigma_k$ and $V^\pm_k$ arise from the Kaluza-Klein reduction of the supergravitons over the $S^3$ and so are on a different footing from the $s_k$ that follow from the reduction of the tensor multiplets. One can encode the R-symmetry indices of each operator in terms the $SU(2)_L\times SU(2)_R$ spinors $A_\alpha$, $\bar{A}_{\dot\alpha}$ or equivalently, to emphasise the higher dimensional origin of the SCFT$_2$ primaries, in terms of an $SO(4)$ vector $t_\mu\equiv  \sigma^\mu_{\alpha \dot\alpha}A^\alpha \bar{A}^{\dot\alpha} $ satisfying\footnote{As usual $\sigma^\mu=(\vec{\sigma},i\, 1_{2\times 2})$, $\bar\sigma^\mu=(\vec{\sigma},-i\, 1_{2\times 2})$ are the chiral blocks of the 4D gamma matrices written in terms of Pauli matrices $\vec{\sigma}$ and the identity.} $t^2=0$:
\begin{equation}
  \label{eq:Ocft}
  \begin{aligned}
   s_k(z_i,\bar{z}_i;t_i) = & ~t_{i\,\mu_1} \ldots t_{i\,\mu_{k}} s_k^{\mu_1 \ldots \mu_{k}}(z_i,\bar{z}_i)\;,\\
    \sigma_k(z_i,\bar{z}_i;t_i) = & ~A_{i\,\alpha} \bar{A}_{i\,\dot \alpha} t_{i\,\mu_1} \ldots t_{i\,\mu_{k-1}} \sigma_k^{\alpha \,\dot{\alpha}_,\mu_1 \ldots \mu_{k-1}}(z_i,\bar{z}_i)\;,
    \\
    V^+_k(z_i,\bar{z}_i;t_i) = & ~A_{i\,\alpha} {A}_{i\,\beta} t_{i\,\mu_1} \ldots t_{i\,\mu_{k}} (V^+_k)^{\alpha \,{\beta}_,\mu_1 \ldots \mu_{k}}(z_i,\bar{z}_i)\;,
  \end{aligned}
\end{equation}
with $V^-_k$ written in a similar way in terms of the bilinear $\bar{A}_{i\,\dot\alpha} \bar{A}_{i\,\dot\beta}$. In the expression for $\sigma_k$ we wrote one of the $SO(4)$ vector indices $\mu_k$ in terms of the bilinear ($\alpha, \dot{\alpha}$). The reason is that, as we will see, it is convenient to consider the descendants obtained by acting on each CPO $O_k(z_i,\bar{z}_i;A_i,\bar{A}_i)$ with an appropriate combination of the supercharges\footnote{The $SU(2)$ indices $\hat{A}$, $\hat{B}$ label an outer isomorphism of the algebra. We will not need to specify the precise form of this supercharges.} $G^{\alpha \hat{A}} \tilde{G}^{\dot{\alpha} \hat{B}}$ yielding a superdescendant $B_k$ in the R-symmetry representation $j_B=j_O-1/2$, $\bar{j}_B=\bar{j}_O-1/2$. The lowest Kaluza-Klein mode of each of these superdescendants will be characterised just by the $A$, $\bar{A}$ without any $t$'s.

Let us start from the simple case of the 4-point correlators involving just the CPOs $s_{k_i}$~\cite{Giusto:2018ovt,Rastelli:2019gtj,Giusto:2019pxc}. For the lowest possible value $k=1$ the superdescendant is a scalar of the R-symmetry group and has spin zero, so it is naturally related to a 6D scalar field in a supergravity tensor multiplet and, as usual, the higher values of $k$ arise from the Kaluza-Klein reduction of the same field with a $S^3$ spherical harmonics of level $k-1$. We saw that, in the flat-space limit, the truncated amplitude $\tilde{A}_{\rm ten}$ in \eqref{eq:Aten} is identical to the tree-level 4-point  correlator of a scalar $\phi^3$ theory in 6D. Thus, if this hidden conformal symmetry holds also in AdS, then it is natural to expect that all CFT$_2$ 4-point  correlators among $s_{k_i}$ are related to a single CFT$_6$ correlator with four 6D {\em scalar} primaries. To show that this is indeed the case, we parametrise the connected tree-level supergravity contribution to the correlator as follows\footnote{For concreteness we are working with the conventions of ``case I'' of Eq.~(2.8) of~\cite{Rastelli:2019gtj}; the final results do not depend on this choice.}  
\begin{equation}
  \label{eq:cft2corr}
  \langle O_{k_1} O_{k_2} O_{k_3} O_{k_4} \rangle^{(1)} = \left(\frac{|\zeta_{13}|^{k_{21}+k_{43}}\,|\zeta_{23}|^{-k_{21}+k_{43}}}{|\zeta_{12}|^{k_1+k_2+k_{43}}|\zeta_{34}|^{2k_4}}\right) \left[{\mathcal{G}}_{\{k_i\}}^{(0)} + \left|\frac{1-\alpha_c z}{1-\alpha_c}\right|^2 \, \left(\widetilde{\mathcal{G}}_{\{k_i\}}+\widetilde{\mathcal{G}}_{\{k_i\}}^{(0)} \right)\right]\;,
\end{equation}
where $k_{ij}=k_i-k_j$, $z_{ij}=z_i-z_j$, $t_{ij} = (t_i-t_j)^2$, $A^i \cdot A^j = A^i_1 A^j_2 - A^i_2 A^j_1$, 
\begin{equation} \label{eq:zalpha}
    |\zeta_{ij}|^2 =\frac{|z_{ij}|^2}{t_{ij}^2}\;,\quad
    z = \frac{z_{14} z_{23}}{z_{13} z_{24}}\,,\quad  
    \alpha_c= \frac{A^1\cdot A^3\,A^2\cdot A^4}{A^1\cdot A^4\,A^2\cdot A^3} \;.
\end{equation}
The superconformal algebra requires that the functions $\widetilde{\mathcal{G}}_{\{k_i\}}$ and $\widetilde{\mathcal{G}}_{\{k_i\}}^{(0)}$ be regular when $\alpha_c\to 1/z$ or $\bar{\alpha}_c \to 1/\bar{z}$ and that $\mathcal{G}^{(0)}_{\{k_i\}}$ become a holomorphic function of $z$ and $\alpha_c$ when $\bar\alpha_c\to 1/\bar{z}$~\cite{Rastelli:2019gtj}. This last condition can be satisfied by taking $\mathcal{G}^{(0)}_{\{k_i\}}$ to be a polynomial in the variables $\sigma\, U$ and $\tau\,U\,V^{-1}$, where $\sigma \equiv \frac{|\alpha_c|^2}{|1-\alpha_c|^2}$, $\tau \equiv \frac{1}{|1-\alpha_c|^2}$, $U\equiv |1-z|^2$, $V \equiv |z|^2$; the order of the polynomial is finite and depends on the $k_i$'s. The split between $\widetilde{\mathcal{G}}_{\{k_i\}}$ and $\widetilde{\mathcal{G}}_{\{k_i\}}^{(0)}$ is required, in general, to single-out the part of the correlator that is encoded in the 6D CFT correlator. It turns out that $\widetilde{\mathcal{G}}_{\{k_i\}}^{(0)}$ can be taken to be a finite-order polynomial in the variables $\sigma\, U$, $\tau\,U\,V^{-1}$ and {\it also} $V^{-1}$. Given the ``dynamical'' part of the correlator $\widetilde{\mathcal{G}}_{\{k_i\}}$, the {\it finite} set of coefficients that are needed to reconstruct $\mathcal{G}^{(0)}_{\{k_i\}}$ and $\widetilde{\mathcal{G}}_{\{k_i\}}^{(0)}$ can be fixed by imposing basic consistency requirements on the OPE in the various channels, like the vanishing of the extremal three-point functions.

With these choices, all $\widetilde{\mathcal{G}}_{\{k_i\}}$'s can be repackaged in a single scalar CFT$_6$ correlator
\begin{equation}
  \label{eq:phi6d}
{\cal C}(Z_i) = \langle \phi(Z_1) \phi(Z_2) \phi(Z_3) \phi(Z_4)\rangle = \frac{f(Z)}{|Z_{12}|^4 |Z_{34}|^4} \, ,
\end{equation}
where $Z_i=(z_i,\bar{z}_i,t^\mu_i)$ are 6D coordinates. Here we took the conformal weight of $\phi$ to be $\Delta_\phi=2$ and parametrised the result in terms of   a single function of the 6D cross ratio $Z$ defined in a similar way to the 2D case~\eqref{eq:zalpha}. The relation between the 6D and 2D CFT correlator is~\cite{Caron-Huot:2018kta,Rastelli:2019gtj}
\begin{equation}
  \label{eq:26}
  \left(\frac{|\zeta_{13}|^{k_{21}+k_{43}}\,|\zeta_{23}|^{-k_{21}+k_{43}}}{|\zeta_{12}|^{k_1+k_2+k_{43}}|\zeta_{34}|^{2k_4}}\right) \, \widetilde{\mathcal{G}}_{\{k_i\}} = c_{\{k_i\}}\,t_{12}^2 t_{34}^2 |z_{13}|^2 |z_{24}|^2 {\cal C}(Z_i)\;,
\end{equation}
where the identity should be interpreted in a Taylor-expanded way by matching the terms with the same number of each $t_i$. The numerical coefficients $c_{\{k_i\}}$ are determined in such a way that $\widetilde{\mathcal{G}}_{\{k_i\}}$ gives the correlator of normalised operators. The function $f(Z)$ can be determined by imposing that~\eqref{eq:26} holds for the correlator $\widetilde{\cal G}_{1111}$ between operators in the lowest Kaluza Klein mode $k_i=1$, using its explicit form found\footnote{See~\cite{Giusto:2018ovt} for our conventions on the functions $\hat{D}$ and how they are related to the Bloch-Wigner dilogarithm and 4-point contact integral in AdS.} in~\cite{Giusto:2018ovt}:
\begin{equation}
  \label{eq:phi6dc}
f(Z)= \frac{2}{\pi} (1-Z)^4 (\delta_{f_1 f_2} \delta_{f_3 f_4}\hat{D}_{1122}(Z)+ \delta_{f_1 f_3} \delta_{f_2 f_4} \hat{D}_{1212}(Z) + \delta_{f_1 f_4} \delta_{f_2 f_3} \hat{D}_{2112}(Z))\;.
\end{equation}
The relation~\eqref{eq:26} was checked in~\cite{Rastelli:2019gtj} for $s_k$ correlators up to $k=3,4$ and it can be shown~\cite{GRTW} to reproduce the Mellin space results of~\cite{Giusto:2019pxc} for all $k$.

\section{The CFT$_6$ 4-point correlator}
\label{sec:cft6}

In order to generalise the approach of the previous section to the full theory, one needs to look for new 6D primaries. Actually a single primary should encode all the remaining CPOs since the fields $\sigma_k$ and $V^\pm_k$ have the same higher dimensional origin from the gravity multiplet. The intuition developed so far is that the 6D primary should be more directly related to the superdescendants. This is also supported by the fact that in 2D/6D relation~\eqref{eq:26}, the combination appearing on the CFT$_2$ side is $\tilde{\cal G}$ without the factor of $|1-\alpha_c z|^2/|1-\alpha_c|^2$ which naturally follows form the Ward identity relating the correlator between the CPOs $O$ and the one between the superdescendants $G\tilde{G} O$~\cite{Giusto:2019pxc}. Notice that this explains the choice of the conformal weight $\Delta_\phi=2$ made above: the quantum numbers of the 6D primary, including the conformal weight, can be read from the lowest Kaluza Klein of the 2D superdescendant: since $B_{s_1}=G\tilde{G} s_1$ has $h=\bar{h}=1$ and $j=\bar{j}=0$, it is natural to relate it to a scalar 6D primary with $\Delta_\phi=2$.

The same argument leads to a superdescendant for $\sigma_2$ with $h=\bar{h}=3/2$ and $j=\bar{j}=1/2$, to a superdescendant for $V^+_1$ with $(h,\bar{h})=(2,1)$ and $(j,\bar{j})=(1,0)$, and similarly with $h\leftrightarrow \bar{h}$, $j\leftrightarrow \bar{j}$ for $V^-_1$. We then have ten degrees of freedom (four from the representation $j=\bar{j}=1/2$ and three each from those with $j=1$ or $\bar{j}=1$) which need to be encoded by a single 6D primary. Then it cannot obviously be a scalar. By identifying the space-time and the R-symmetry groups of the CFT$_2$ with the decomposition $SO(2,2)\times SO(4) \subset SO(2,6)$ of the 6D conformal group, we deduce the the 6D primary should contain a vector of $SO(4)$ which is a Lorentz scalar and a 2-form of $SO(4)$ which is a $SO(1,1)$ 1-form in space-time. Actually the latter should split in two irreps with the self-duality of the $SO(4)$ and $SO(1,1)$ parts linked together, for instance taking them to be both self-dual to describe $V^+_1$ or both anti-self-dual to describe $V^-_1$. In summary, this suggests to consider a 6D primary field which is a self-dual 3-form $w$ with $\Delta_w=3$. Notice that such primary has ten of degrees of freedom as required by the counting above.

It is now clear how to make the hidden 6D symmetry of our general CFT$_2$ correlators manifest. One should first start from the 6D correlator involving the appropriate number of scalar fields $\phi$ and self-dual 3-forms $w$
\begin{equation}
  \label{eq:wsd}
  w_{m_1 m_2 m_3} = -\frac{i}{3!} \epsilon_{m_1 m_2 m_3 m_4 m_5 m_6} w^{m_4 m_5 m_6}\,,
\end{equation}
and parametrise its most general expression in terms of arbitrary functions of the 6D cross ratio $Z$. These function can be determined by using some explicit CFT$_2$ data from correlators involving low Kaluza-Klein modes. Finally the generic CFT$_2$ correlators can be extracted again from~\eqref{eq:26} where now the appropriate 6D correlator ${\cal C}(Z_i)$ is used. 

As an example, here we will work out the 6D correlator involving two scalars and two 3-forms $\langle \phi(Z_1) \phi(Z_2) w^{(3)}(Z_3) w^{(4)}(Z_4)\rangle$. This should capture all CFT$_2$ correlators with two $s_k$ CPOs and two CPOs in the $\sigma_k$ or $V^\pm_k$ multiplets. In order to write the most general 6D correlator in this case we proceed in two steps. First, we need to count the number of independent functions present in this result which can be easily done by following~\cite{Kravchuk:2016qvl}. The logic is to fix a conformal frame where two operators are in a plane: for instance we can take the scalars to be in the CFT$_2$ directions. Then we take the polarizations $w_{m_1 m_2 m_3}$ of the 3-forms and decompose the 6D indices in the CFT$_2$ directions $a=1,2$ and the remaining $SO(4)$ directions $\mu=1,\ldots, 4$. There is an independent function in the general expression of the correlator for each $SO(4)$-invariant combination we can construct from the 3-form polarizations; of course in the counting we need to impose the 6D self-duality constraint, which means we can focus on the $SO(4)$ scalars obtained from the independent components $w_{\mu\nu\rho}$ and $w_{\mu\nu a}$. For the case of two scalars and two 3-forms mentioned above we have four structures linear in both $w^{(3)}$ and $w^{(4)}$.

Then we need an explicit expression for each of the four independent structures and for this, it is convenient to follow the embedding formalism~\cite{Costa:2011mg,Costa:2014rya}. We introduce 8D coordinates $P^M_i=(P^+_i,P^-_i,Z^m_i)=(1,|z_i|^2 + t_i^2,z^a_i,t^\mu_i)$, with a metric $(P_i,P_j)=P^+_i P^-_j + P^-_i P^+_j - 2 Z^m_i Z^m_j$. Similarly we promote $w^{(i)}$ to an antisymmetric tensor in 8D $W^{(i)}_{M_1 M_2 M_3}$ which is transverse $W^{(i)}_{M_1 M_2 N}P^N_{(i)}=0$ and whose pull-back in 6D agrees, of course, with the original polarization $w^{(i)}_{m_1 m_2 m_3}$. It is possible to consistently impose $W^{(i)}_{MN+}=0$ and the remaining components of the 3-form in the embedding space read
\begin{equation}\label{eq:projectioninv}
W^{(i)}_{mn-}=-\frac{Z_i^r}{Z_i^{2}} W^{(i)}_{mnr} \,,~~
W^{(i)}_{m n r}=w^{(i)}_{mnr}-\frac{2 Z_i^s}{Z_i^{2}} (Z_{i\, m} w^{(i)}_{n r s}- Z_{i\,_n} w^{(i)}_{m r s}+Z_{i\,r} w^{(i)}_{m n s})\,.
\end{equation}
If we restrict ourselves to the correlators containing only $\sigma$'s we should take 
\begin{equation}\label{eq:Wsigma}
w^{(i)}_{\mu\nu\rho} = \epsilon_{\mu\nu\rho\sigma} \,\hat{t}_{i\,\sigma}\,,\quad w^{(i)}_{ab\mu}= \,i\, \epsilon_{ab} \,\hat{t}_{i\,\mu}\,,
\end{equation}
with all other components of $w_{mnr}$ set to zero. In order to preserve the 6D conformal invariance one should take arbitrary, but constant vectors $\hat{t}_i$. For the purposes of making contact with the CFT$_2$ correlators, we break the full 6D symmetry to $SO(2,2)\times SO(4)$ by identifying $\hat{t}_i$ with $t_i$, the 4D part of the position $Z_i$: $\hat{t}_i\equiv t_i$. If we are to describe the operators $V^{\pm}$, we conjecture that one should take $w_{\mu\nu\rho}=w_{\mu a b}=0$ and $w_{\mu\nu a}$ as follows 
\begin{equation}\label{eq:WV}
  w^{(i) \pm}_{\mu\nu 1} = \hat{t}^\pm_{i\,\mu\nu}\,,~~ w^{(i) \pm}_{\mu\nu 2} = \pm i\, \hat{t}^{\pm}_{i\,\mu\nu}\,,~~
\hat{t}^{\pm}_{i\,\mu\nu}=\pm \frac{1}{2} \epsilon_{\mu\nu\rho\sigma} \hat{t}^{\pm}_{i\,\rho\sigma}\,,~~  \hat{t}^{\pm}_{i\,\mu\nu} t_i^{\mu}=0\,.
\end{equation}
As before $\hat{t}_{i\,\mu\nu}$ should be a constant polarization to preserve the 6D invariance, but here we link it to the 4D part $t_i$ of $Z_i$: $\hat{t}^+_{i\,\mu\nu}= {t}^+_{i\,\mu\nu}\equiv \bar{A}_{i\,\dot \alpha} \bar{A}_{i\,\dot \beta} \bar\sigma^{\dot \alpha \dot\beta}_{\mu\nu}$, $\hat{t}^{-}_{i\,\mu\nu}= {t}^{-}_{i\,\mu\nu} \equiv {A}_{i\,\alpha} {A}_{i\,\beta} \sigma^{\alpha \beta}_{\mu\nu}$. This automatically solves the constraints in~\eqref{eq:WV}}. Of the four possible 6D conformal structures for $\langle \phi(Z_1) \phi(Z_2) w^{(3)}(Z_3) w^{(4)}(Z_4)\rangle$, it turns out that only three of them are independent when $\hat{t}_i$ is identified with $t_i$ in the polarizations~\eqref{eq:Wsigma} and~\eqref{eq:WV}; they can be written in terms of the following expressions in embedding space
\begin{align}\label{eq:Sij}
{\cal S}^{ij} \,\, = & - \frac{4}{ (P^{(1)}, P^{(2)})}\Big[(W^{(3)}_{MNA} W_{(4)}^{MNB} P^{(i)}_A P^{(j)}_B)(P^{(3)}, P^{(4)})\\ \nonumber
&- (W^{(3)}_{MNA} W_{(4)}^{MNB} P^{A}_{(4)} P^{(j)}_B)(P^{(i)}, P^{(3)})- (W^{(3)}_{MNA} W_{(4)}^{MNB} P^{(3)}_B P^{A\,(i)})(P^{(j)}, P^{(4)})\\ \nonumber
&-2 (W^{(3)}_{MPA} W_{(4)}^{MQB} P^{(3)}_Q P^{P\,(4)} P^{A\,(i)} P^{(j)}_B)+\frac{1}{3} (W^{(3)}_{MNP} W_{(4)}^{MNP}) (P^{(i)}, P^{(3)}) (P^{(j)}, P^{(4)})\Big]\,,
\end{align}
and
\begin{align}\label{eq:Tij}
{\cal T}^{ij} \,\, = - \frac{ 16\,i\, \epsilon^{M_1\ldots M_8}}{(P^{(1)}, P^{(2)})^2} \Bigl [ &W^{(3)}_{M_1 M_2 P} W^{(4)}_{M_3 M_4 Q} P^{(3)}_{M_5}  P^{(4)}_{M_6} P^{(1)}_{M_7} P^{(2)}_{M_8} \,P_{(i)}^{P} P_{(j)}^{Q} \\ \nonumber
&-\frac{1}{3} W^{(3)}_{M_1 M_2 M_3} W^{(4)}_{M_4 M_5 Q} P^{(4)}_{M_6} P^{(1)}_{M_7} P^{(2)}_{M_8} \,P_{(j)}^{Q} (P^{(i)}, P^{(3)})\\ \nonumber
&+\frac{1}{3} W^{(3)}_{M_1 M_2 P} W^{(4)}_{M_3 M_4 M_5} P^{(3)}_{M_6} P^{(1)}_{M_7} P^{(2)}_{M_8} \,P_{(i)}^{P} (P^{(j)}, P^{(4)})\\ \nonumber
&+\frac{1}{9} W^{(3)}_{M_1 M_2 M_3} W^{(4)}_{M_4 M_5 M_6} P^{(1)}_{M_7} P^{(2)}_{M_8} \, (P^{(i)}, P^{(3)})(P^{(j)}, P^{(4)})\Bigr ]\,,
\end{align}
where we can choose $(i,j)=(1,2)$ or  $(i,j)=(2,1)$. The symmetric and antisymmetric parts of ${\cal S}^{ij}$, and the antisymmetric part of ${\cal T}^{ij}$ yield independent structures, so we can parametrise the 6D correlator in terms of three functions $f_i(Z)$ as follows
\begin{equation}
  \label{eq:phi2w26d}
 \langle \phi(Z_1) \phi(Z_2) w^{(3)}(Z_3) w^{(4)}(Z_4)\rangle = \frac{1}{|Z_{12}|^4 |Z_{34}|^8} \left[f_1(Z)\, {\cal S}^{(12)} - f_2(Z) \, {\cal S}^{[12]} + f_3(Z)\, {\cal T}^{[12]} \right]\,.
\end{equation}
         
\section{A new recursion relation}
\label{sec:sssigmasigma}

In this section we focus on the case where the 3-form 6D primary describes the CFT$_2$ CPO $\sigma_k$ in~\eqref{eq:Ocft}. As a first step, we use~\eqref{eq:Wsigma} in~\eqref{eq:phi2w26d} in order to write explicitly ${\cal S}$ and ${\cal T}$ in terms of the 2D coordinates and the R-symmetry variables obtaining
\begin{equation}
\begin{aligned} \label{eq:s2sigma2}
{\cal C}(Z_i)= & \langle \phi(Z_1) \phi(Z_2) w^{(3)}(Z_3) w^{(4)}(Z_4)\rangle  =\frac{1}{|Z_{12}|^4 |Z_{34}|^8}\Bigl\{f_1(Z)\, t_{34}^2 |z_{34}|^2\\
&+ f_2(Z) \Bigl[  \left(t_{14}^2 t_{23}^2 - t_{13}^2 t_{24}^2 \right) \frac{|z_{34}|^2}{|Z_{12}|^2} + t_{34}^2 \frac{|z_{14}|^2 |z_{23}|^2-|z_{13}|^2 |z_{24}|^2}{|Z_{12}|^2}\Bigr]\\
 &+ f_3(Z) \Bigl[ \frac{(t_{13}^2 t_{24}^2+t_{14}^2t_{23}^2)|z_{34}|^2+t_{34}^2 (|z_{14}|^2 |z_{23}|^2+|z_{13}|^2|z_{24}|^2)}{2 |Z_{12}|^2}\\
  & - \frac{t_{12}^2 t_{34}^2 |z_{13}|^2 |z_{24}|^2 (z+\bar z)}{|Z_{12}|^4} - \frac{t_{13}^2 t_{14}^2 |z_{23}|^2 |z_{24}|^2 +t_{23}^2 t_{24}^2 |z_{13}|^2 |z_{14}|^2 }{|Z_{12}|^4} \\
 & - \frac{t_{13}^2 t_{24}^2 (|z_{12}|^2 |z_{34}|^2- |z_{13}|^2 |z_{24}|^2)+t_{14}^2 t_{23}^2 (|z_{12}|^2 |z_{34}|^2- |z_{14}|^2 |z_{23}|^2)}{|Z_{12}|^4} \\
 & - 4\, \epsilon_{\mu_1 \mu_2 \mu_3 \mu_4} t_1^{\mu_1} t_2^{\mu_2} t_3^{\mu_3} t_4^{\mu_4}\frac{|z_{13}|^2 |z_{24}|^2}{|Z_{12}|^4} (z-\bar z) \Bigr] \Bigr\}\;.
 \end{aligned}
\end{equation}
Then, as for the case discussed in section~\ref{sec:hints}, we should match the general expression~\eqref{eq:s2sigma2} with some explicit results for $\langle s_{k_1} s_{k_2} \sigma_{k_3} \sigma_{k_4} \rangle^{(1)}$ in the CFT$_2$. The matching is done again by using~\eqref{eq:26}, but now using~\eqref{eq:s2sigma2} for ${\cal C}(Z_i)$. It is, of course, convenient to use the correlators involving the lowest Kaluza-Klein modes, so we will use the functions $\widetilde{\mathcal{G}}_{1122}$ and $\widetilde{\mathcal{G}}_{2222}$ as defined in~\eqref{eq:cft2corr}; notice that while $\widetilde{\mathcal{G}}_{1122}$ does not depend on $\alpha_c$, $\widetilde{\mathcal{G}}_{2222}$ contains four different $\alpha_c$-dependent structures, proportional to $1$, $\sigma$, $\tau$ and $(\alpha_c-\bar{\alpha}_c)/|1-\alpha_c|^2$, so these correlators already over-constrain the problem. The structure $(\alpha_c-\bar{\alpha}_c)/|1-\alpha_c|^2$ cannot be written in terms of scalar products of the $t_i$'s and, hence, it cannot arise from a 6D correlator between two scalars and two symmetric tensors of any spin. One can use this term to fix $f_3$ since the only structure that can yield such combination is the one in the last line of~\eqref{eq:s2sigma2}. By using the results in~\cite{GRTW} we have
\begin{align}\label{eq:f3}
f_3(z) =& \frac{16}{3\,c_{2222}}\,|1-z|^8 \left( \frac{1}{2}\,\hat D_{2123}-\frac{1}{3}\,\hat D_{2233}-\frac{1}{3}\,\hat D_{3223}\right)\,.
\end{align}
Then one can focus on the term proportional to $\tau$ and determine the function $f_2$
\begin{equation}\label{eq:f2}
f_2(z)=\frac{8}{3\,c_{2222}}\,|1-z|^6 \left[ \frac{1}{2} (\hat D_{1223}-\hat D_{2123}) +\frac{1}{3}  (\hat D_{2323}-\hat D_{3223}) \right]\,.
\end{equation}
The term proportional to $\sigma$ in $\widetilde{\mathcal{G}}_{2222}$ involves the same functions $f_{2}$, $f_{3}$ and provides a first consistency check of~\eqref{eq:s2sigma2}. Finally from $\widetilde{\mathcal{G}}_{1122}$ and the $\tau$ and $\sigma$ components of $\widetilde{\mathcal{G}}_{2222}$ we deduce
\begin{equation}\label{eq:f1}
f_1(z) = -\frac{16}{c_{2222}}\,|1-z|^8 \hat D_{1144}\,,~~ c_{2222}=-2 c_{1122}\;.
\end{equation}
A further consistency check comes by using the functions $f_i$ to compute the $\alpha_c$-independent component of $\widetilde{\mathcal{G}}_{2222}$. It is now possible to expand~\eqref{eq:26} with ${\cal C}(Z_i)$ given by~\eqref{eq:s2sigma2} and~\eqref{eq:f3}--\eqref{eq:f1} to obtain predictions for correlators with arbitrary weights. We checked~\cite{GRTW} that the results are consistent with the explicit correlators of several different weights $(k_1,k_2,k_3,k_4)$: $(1,1,l,l)$ for arbitrary $l$, $(3,1,2,2)$, $(2,2,3,3)$, and $(3,3,2,2)$. 

%%%%%%%%%%%%%%%%%%%%%%%%%%%%%%%%%%%%%%%%%%%%%%%%%%%%%%%%%%%%
\section*{Acknowledgements}

We would like to thank Alessandro Bombini and Andrea Galliani for collaboration at an early stage of this work and Francesco Aprile, Nejc \v{C}eplak and Marcel Hughes for discussions. This work was supported in part by the Science and Technology Facilities Council (STFC) Consolidated Grant ST/P000754/1 {\it String theory, gauge theory \& duality} and by the MIUR-PRIN contract 2017CC72MK003. C.W. is supported by a Royal Society University Research Fellowship No. UF160350.

\providecommand{\href}[2]{#2}\begingroup\raggedright\endgroup

%\bibliographystyle{utphys}      % (uses file "utphys.bst")
%\bibliography{microstates2}              % expects file "micrpdostates.bib"

\begin{thebibliography}{10}

\bibitem{Witten:1998qj}
E.~Witten, ``{Anti-de Sitter space and holography},'' {\em Adv. Theor. Math.
  Phys.} {\bf 2} (1998) 253--291,
\href{http://arXiv.org/abs/hep-th/9802150}{{\tt hep-th/9802150}}.
%%CITATION = HEP-TH/9802150;%%.

\bibitem{Penedones:2010ue}
J.~Penedones, ``{Writing CFT correlation functions as AdS scattering
  amplitudes},'' {\em JHEP} {\bf 03} (2011) 025,
\href{http://arXiv.org/abs/1011.1485}{{\tt 1011.1485}}.
%%CITATION = ARXIV:1011.1485;%%.

\bibitem{Fitzpatrick:2011ia}
A.~Fitzpatrick, J.~Kaplan, J.~Penedones, S.~Raju, and B.~C. van Rees, ``{A
  Natural Language for AdS/CFT Correlators},'' {\em JHEP} {\bf 11} (2011) 095,
  \href{http://arXiv.org/abs/1107.1499}{{\tt 1107.1499}}.

\bibitem{Rastelli:2016nze}
L.~Rastelli and X.~Zhou, ``{Mellin amplitudes for $AdS_5\times S^5$},'' {\em
  Phys. Rev. Lett.} {\bf 118} (2017), no.~9, 091602,
\href{http://arXiv.org/abs/1608.06624}{{\tt 1608.06624}}.
%%CITATION = ARXIV:1608.06624;%%.

\bibitem{Rastelli:2017udc}
L.~Rastelli and X.~Zhou, ``{How to Succeed at Holographic Correlators Without
  Really Trying},'' {\em JHEP} {\bf 04} (2018) 014,
\href{http://arXiv.org/abs/1710.05923}{{\tt 1710.05923}}.
%%CITATION = ARXIV:1710.05923;%%.

\bibitem{Alday:2016njk}
L.~F. Alday, ``{Large Spin Perturbation Theory for Conformal Field Theories},''
  {\em Phys. Rev. Lett.} {\bf 119} (2017), no.~11, 111601,
\href{http://arXiv.org/abs/1611.01500}{{\tt 1611.01500}}.
%%CITATION = ARXIV:1611.01500;%%.

\bibitem{Caron-Huot:2017vep}
S.~Caron-Huot,
``{Analyticity in Spin in Conformal Theories},''
{\em JHEP} {\bf 09} (2017) 078, 
\href{http://arXiv.org/abs/1703.00278}{{\tt 1703.00278}}.
%179 citations counted in INSPIRE as of 17 May 2020

\bibitem{Alday:2017vkk}
L.~F. Alday and S.~Caron-Huot, ``{Gravitational S-matrix from CFT dispersion
  relations},'' {\em JHEP} {\bf 12} (2018) 017,
\href{http://arXiv.org/abs/1711.02031}{{\tt 1711.02031}}.
%%CITATION = ARXIV:1711.02031;%%.

\bibitem{Galliani:2017jlg}
A.~Galliani, S.~Giusto, and R.~Russo, ``{Holographic 4-point correlators with
  heavy states},'' {\em JHEP} {\bf 10} (2017) 040,
\href{http://arXiv.org/abs/1705.09250}{{\tt 1705.09250}}.
%%CITATION = ARXIV:1705.09250;%%.

\bibitem{Bombini:2017sge}
A.~Bombini, A.~Galliani, S.~Giusto, E.~Moscato, and R.~Russo, ``{Unitary
  4-point correlators from classical geometries},'' {\em Eur. Phys. J.} {\bf
  C78} (2018), no.~1, 8,
\href{http://arXiv.org/abs/1710.06820}{{\tt 1710.06820}}.
%%CITATION = ARXIV:1710.06820;%%.

\bibitem{Aprile:2017xsp}
F.~Aprile, J.~M. Drummond, P.~Heslop, and H.~Paul, ``{Unmixing Supergravity},''
  {\em JHEP} {\bf 02} (2018) 133,
\href{http://arXiv.org/abs/1706.08456}{{\tt 1706.08456}}.
%%CITATION = ARXIV:1706.08456;%%.

\bibitem{Aprile:2018efk}
F.~Aprile, J.~Drummond, P.~Heslop, and H.~Paul, ``{Double-trace spectrum of
  $N=4$ supersymmetric Yang-Mills theory at strong coupling},'' {\em Phys.
  Rev.} {\bf D98} (2018), no.~12, 126008,
\href{http://arXiv.org/abs/1802.06889}{{\tt 1802.06889}}.
%%CITATION = ARXIV:1802.06889;%%.

\bibitem{Caron-Huot:2018kta}
S.~Caron-Huot and A.-K. Trinh, ``{All tree-level correlators in ${\rm AdS}_{5}
  \times {\rm S}_{5}$ supergravity: hidden ten-dimensional conformal
  symmetry},'' {\em JHEP} {\bf 01} (2019) 196,
  \href{http://arXiv.org/abs/1809.09173}{{\tt 1809.09173}}.

\bibitem{Rastelli:2019gtj}
L.~Rastelli, K.~Roumpedakis, and X.~Zhou, ``{$\mathbf{AdS_3\times S^3}$
  Tree-Level Correlators: Hidden Six-Dimensional Conformal Symmetry},'' {\em
  JHEP} {\bf 10} (2019) 140,
\href{http://arXiv.org/abs/1905.11983}{{\tt 1905.11983}}.
%%CITATION = ARXIV:1905.11983;%%.

\bibitem{Giusto:2019pxc}
S.~Giusto, R.~Russo, A.~Tyukov, and C.~Wen, ``{Holographic correlators in
  AdS$_3$ without Witten diagrams},'' {\em JHEP} {\bf 09} (2019) 030,
\href{http://arXiv.org/abs/1905.12314}{{\tt 1905.12314}}.
%%CITATION = ARXIV:1905.12314;%%.

\bibitem{Zhou:2018ofp}
X.~Zhou, ``{On Mellin Amplitudes in SCFTs with Eight Supercharges},'' {\em
  JHEP} {\bf 07} (2018) 147, \href{http://arXiv.org/abs/1804.02397}{{\tt
  1804.02397}}.

\bibitem{Heydeman:2018dje}
M.~Heydeman, J.~H. Schwarz, C.~Wen, and S.-Q. Zhang, ``{All Tree Amplitudes of
  6D $(2,0)$ Supergravity: Interacting Tensor Multiplets and the $K3$ Moduli
  Space},'' {\em Phys. Rev. Lett.} {\bf 122} (2019), no.~11, 111604,
\href{http://arXiv.org/abs/1812.06111}{{\tt 1812.06111}}.
%%CITATION = ARXIV:1812.06111;%%.


\bibitem{Giusto:2018ovt}
S.~Giusto, R.~Russo, and C.~Wen, ``{Holographic correlators in AdS$_{3}$},''
  {\em JHEP} {\bf 03} (2019) 096,
\href{http://arXiv.org/abs/1812.06479}{{\tt 1812.06479}}.
%%CITATION = ARXIV:1812.06479;%%.

\bibitem{GRTW}
S.~Giusto, R.~Russo, A.~Tyukov, and C.~Wen, ``In preparation,''.

\bibitem{Alday:2017xua}
L.~F. Alday and A.~Bissi, ``{Loop Corrections to Supergravity on $AdS_5 \times
  S^5$},'' {\em Phys. Rev. Lett.} {\bf 119} (2017), no.~17, 171601,
\href{http://arXiv.org/abs/1706.02388}{{\tt 1706.02388}}.
%%CITATION = ARXIV:1706.02388;%%.

\bibitem{Aprile:2017bgs}
F.~Aprile, J.~M. Drummond, P.~Heslop, and H.~Paul, ``{Quantum Gravity from
  Conformal Field Theory},'' {\em JHEP} {\bf 01} (2018) 035,
\href{http://arXiv.org/abs/1706.02822}{{\tt 1706.02822}}.
%%CITATION = ARXIV:1706.02822;%%.

\bibitem{Aprile:2017qoy}
F.~Aprile, J.~Drummond, P.~Heslop, and H.~Paul, ``{Loop corrections for
  Kaluza-Klein AdS amplitudes},'' {\em JHEP} {\bf 05} (2018) 056,
  \href{http://arXiv.org/abs/1711.03903}{{\tt 1711.03903}}.

\bibitem{Alday:2018pdi}
L.~F. Alday, A.~Bissi, and E.~Perlmutter, ``{Genus-One String Amplitudes from
  Conformal Field Theory},'' {\em JHEP} {\bf 06} (2019) 010,
  \href{http://arXiv.org/abs/1809.10670}{{\tt 1809.10670}}.

\bibitem{Alday:2018kkw}
L.~F. Alday, ``{On Genus-one String Amplitudes on $AdS_5 \times S^5$},''
  \href{http://arXiv.org/abs/1812.11783}{{\tt 1812.11783}}.

\bibitem{Drummond:2019odu}
J.~Drummond, D.~Nandan, H.~Paul, and K.~Rigatos, ``{String corrections to AdS
  amplitudes and the double-trace spectrum of $ \mathcal{N} $ = 4 SYM},'' {\em
  JHEP} {\bf 12} (2019) 173, \href{http://arXiv.org/abs/1907.00992}{{\tt
  1907.00992}}.

\bibitem{Aprile:2019rep}
F.~Aprile, J.~Drummond, P.~Heslop, and H.~Paul, ``{One-loop amplitudes in
  AdS$_{5} \times S^{5}$ supergravity from $ \mathcal{N} $ = 4 SYM at strong
  coupling},'' {\em JHEP} {\bf 03} (2020) 190,
  \href{http://arXiv.org/abs/1912.01047}{{\tt 1912.01047}}.

\bibitem{Alday:2019nin}
L.~F. Alday and X.~Zhou, ``{Simplicity of AdS Supergravity at One Loop},''
  \href{http://arXiv.org/abs/1912.02663}{{\tt 1912.02663}}.

\bibitem{Drummond:2020dwr}
J.~Drummond, H.~Paul, and M.~Santagata, ``{Bootstrapping string theory on
  AdS$_5 \times S^5$},'' \href{http://arXiv.org/abs/2004.07282}{{\tt
  2004.07282}}.

\bibitem{Binder:2019jwn}
D.~J. Binder, S.~M. Chester, S.~S. Pufu, and Y.~Wang, ``{$ \mathcal{N} $ = 4
  Super-Yang-Mills correlators at strong coupling from string theory and
  localization},'' {\em JHEP} {\bf 12} (2019) 119,
  \href{http://arXiv.org/abs/1902.06263}{{\tt 1902.06263}}.

\bibitem{Chester:2019pvm}
S.~M. Chester, ``{Genus-2 holographic correlator on AdS$_{5} \times S^{5}$ from
  localization},'' {\em JHEP} {\bf 04} (2020) 193,
  \href{http://arXiv.org/abs/1908.05247}{{\tt 1908.05247}}.

\bibitem{Chester:2019jas}
S.~M. Chester, M.~B. Green, S.~S. Pufu, Y.~Wang, and C.~Wen, ``{Modular
  Invariance in Superstring Theory From ${\cal N} = 4$ Super-Yang Mills},''
  \href{http://arXiv.org/abs/1912.13365}{{\tt 1912.13365}}.

\bibitem{Chester:2020dja}
S.~M. Chester and S.~S. Pufu, ``{Far Beyond the Planar Limit in
  Strongly-Coupled $\mathcal{N}=4$ SYM},''
  \href{http://arXiv.org/abs/2003.08412}{{\tt 2003.08412}}.

\bibitem{Huang:2010rn}
Y.-t. Huang and A.~E. Lipstein, ``{Amplitudes of 3D and 6D Maximal
  Superconformal Theories in Supertwistor Space},'' {\em JHEP} {\bf 10} (2010)
  007, \href{http://arXiv.org/abs/1004.4735}{{\tt 1004.4735}}.

\bibitem{Kravchuk:2016qvl}
P.~Kravchuk and D.~Simmons-Duffin, ``{Counting Conformal Correlators},'' {\em
  JHEP} {\bf 02} (2018) 096, \href{http://arXiv.org/abs/1612.08987}{{\tt
  1612.08987}}.

\bibitem{Costa:2011mg}
M.~S. Costa, J.~Penedones, D.~Poland, and S.~Rychkov, ``{Spinning Conformal
  Correlators},'' {\em JHEP} {\bf 11} (2011) 071,
\href{http://arXiv.org/abs/1107.3554}{{\tt 1107.3554}}.
%%CITATION = ARXIV:1107.3554;%%.

\bibitem{Costa:2014rya}
M.~S. Costa and T.~Hansen, ``{Conformal correlators of mixed-symmetry
  tensors},'' {\em JHEP} {\bf 02} (2015) 151,
  \href{http://arXiv.org/abs/1411.7351}{{\tt 1411.7351}}.

\end{thebibliography}

\end{document}